\documentclass[showpacs,showkeys]{revtex4}
\usepackage[dvips]{graphicx}
\usepackage{epsfig}
\begin{document}

\title{Asymptotic Hamiltonian reduction \\
       for  the dynamics of a particle on a surface}

\author{V.L. Golo$^1$}
\email{golo@mech.math.msu.su }

\author{D.O. Sinitsyn$^1$}

\affiliation{$^1$ Department of Mechanics and Mathematics \\
     Moscow University \\
     Moscow 119 899   GSP-2, Russia \\  }

\date{November 5, 2005}

\begin{abstract}
    We consider the motion of a particle on a surface which is
    a small perturbation of the standard sphere. One may
    qualitatively describe the motion  by  means of a precessing
    great circle of the sphere. The observation is employed
    to derive a subsidiary Hamiltonian system
    that has the form of equations for the top with a 4-th order
    Hamiltonian, and provides the detailed
    asymptotic description of the particle's motion in terms of
    graphs on the standard sphere.
\end{abstract}

\pacs{1111} \keywords{motion on surface, asymptotic, averaging
method, separatrixe}

\maketitle

    \section{Introduction}
    \label{introduction}

The dynamics of a particle which is allowed to move freely, i.e.
without the action of external forces, on a smooth surface is the
classical problem in analytical dynamics, \cite{Wh}. It is
generally hard to solve. In fact, the four Hamiltonian equations
for orbits on a surface can be reduced to a system of two
Hamiltonian equations by use of the integral of energy and
elimination of time, \cite{Wh}. But the system obtained in this
way may have no further integrals and admit of no exact solutions.
Thus, the usual reduction method, \cite{Wh}, or the momentum map
according to the current terminology, does not work. The problem
looks even more pessimistic if one aims at drawing a picture of
the ensemble of orbits on a surface, for it requires the study of
the general form and disposition of orbits on a surface of general
shape.

In this paper the surface is supposed to be a perturbed standard
sphere.  Using  the perturbation theory we construct a Hamiltonian
system, which enables us to give a fairly detailed picture of the
ensemble of orbits by means of graphs on the standard sphere; the
vertices of the graphs corresponding to orbits which are
asymptotically closed and the edges of the graphs to orbits
joining the almost closed ones.

\begin{figure}
  \begin{center}
    \includegraphics[width = 200bp]{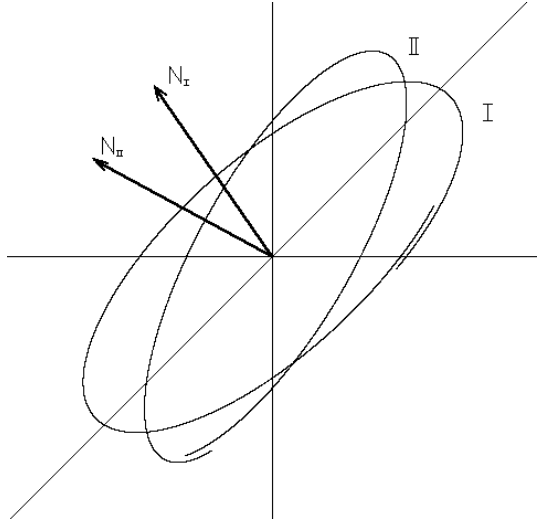}
    \caption{Two coils I, II of an orbit on the surface
     $ \varphi(\vec x) = 0 $;
     vectors $ N_I $ and $ N_{II} $ are the normals to the planes of
     great circles approximating the coils}   \label{fig1}
  \end{center}
\end{figure}

To be specific, the central idea relies on the circumstance that
if the surface does not differ substantially from a sphere, great
circles of the latter may serve a good approximation to orbits of
the particle.  If segments of an orbit  are short enough, one can
visualize it as winding up in coils; loops or rings of the coil
corresponding to great circles of the sphere, see FIG.\ref{fig1}.
Hence approximating the successive rings by great circles, we may
describe the change in the position of the rings by the precession
of a great circle, which in its turn is determined by the normal
vector $\vec L$ of the plane cutting the sphere by the great
circle. To cast this picture in a more quantitative form we may
use the fact that the normal vector $\vec L$ is the angular
momentum of the particle moving on the great circle. Thus, we may
use perturbation theory, i.e. averaging method, for determining
the motion of the angular momentum.

    \section{Averaged equations of motion}
    \label{main_eq}

The equations of motion of a particle on a surface given by the
equation $ \varphi(\vec x) = 0 $ can be cast in the form of the
equation, \cite{Wh}, \cite{R},

  \begin{equation}
    \ddot{\vec x}
    = \lambda \,
    \frac{\partial \varphi}
     {\partial \vec x}   \mbox{.}
    \label{2Newton}
  \end{equation}

The Lagrangian multiplier can be found explicitly , so that the
equation of motion, in the form that does not involve $\lambda$,
reads

  \begin{equation}
    \ddot{\vec x} = - \frac{\dot{\vec x} \cdot
          \displaystyle
          \frac{\partial^2 \varphi}{\partial
                \vec x^2} \cdot
           \, \dot{\vec x}
          }
         {\left( \displaystyle
         \frac{\partial \varphi}{\partial \vec x}
          \right)^2} \,
         \frac{\partial \varphi}{\partial \vec x}    \mbox{.}
    \label{2Newt_last}
  \end{equation}

\noindent In this paper we assume that the equations of the
surface be of the form

$$
  \varphi(\vec x) = \sum_i ( x_i^2  +  \varepsilon_i x_i^4 ) - 1
  \mbox{,}
$$

\noindent where $\varepsilon_i$ are small. Then
equations (\ref{2Newt_last}) read

  \begin{equation}
    \ddot{x_i} = - \frac{
     \sum_j (2 + 12 \varepsilon_j x_j^2) \dot{x_j}^2}
    {
     \sum_j (2 x_j + 4 \varepsilon_j x_j^3) ^2}
       (2 x_i + 4 \varepsilon_i x_i^3)   \mbox{.}
    \label{ConcrNewt}
  \end{equation}

The angular momentum $ \vec L = \vec r \times \vec p.$ verifies
the equations, which follow from (\ref{ConcrNewt})

  \begin{equation}
    \begin{array}{lcl}
    \dot{L_1} = - 4 \, \displaystyle \frac{
     \sum_j (2 + 12 \varepsilon_j x_j^2) \dot{x_j}^2}
       {
     \sum_j (2 x_j + 4 \varepsilon_j x_j^3) ^2} \,
       x_2 x_3 (\varepsilon_3 x_3^2 - \varepsilon_2 x_2^2)
    \vspace{2mm} \\
    \dot{L_2} = - 4 \, \displaystyle \frac{
     \sum_j (2 + 12 \varepsilon_j x_j^2) \dot{x_j}^2}
       {
     \sum_j (2 x_j + 4 \varepsilon_j x_j^3) ^2} \,
       x_3 x_1 (\varepsilon_1 x_1^2 - \varepsilon_3 x_3^2)
    \vspace{2mm} \\
    \dot{L_3} = - 4 \, \displaystyle \frac{
     \sum_j (2 + 12 \varepsilon_j x_j^2) \dot{x_j}^2}
       {
     \sum_j (2 x_j + 4 \varepsilon_j x_j^3) ^2} \,
       x_1 x_2 (\varepsilon_2 x_2^2 - \varepsilon_1 x_1^2)
    \mbox{.}
    \vspace{2mm}
    \end{array}
    \label{ConcrMom}
  \end{equation}

\noindent  The equations given above are exact, in the sense that
they do not involve any approximation and do not use the
$\varepsilon_i$ being small.  Their treatment still needs further
refining, and this will be done with the help of the method of
averaging. Generally, the  approach relies on studying the
evolution equations for integrals of motion of the unperturbed
system, i.e. in our case the normals to the planes of the large
circles, with respect to the basic periodic solution of the
latter. The averaging serves as a filter separating the main
regular part of the solution from the oscillating one caused by
small terms considered as perturbation, see \cite{H}.

We may write the basic equation for the particle's  motion on
the sphere of unit radius in the form

$$
    \vec x = cos(\omega t + \theta) \vec e_1
       + sin(\omega t + \theta) \vec e_2
$$

\noindent where vectors $\vec e_1, \vec e_2, \vec e_3$ can be
conveniently taken in the form

$$
\begin{array}{lcl}
  \vec e_1 = \displaystyle \frac{1}{\sqrt{L_2^2 + L_3^2}}(0, L_3, -L_2)
    \vspace{2mm} \\
  \vec e_2 = \displaystyle \frac{1}{L \sqrt{L_2^2 + L_3^2}}
    (-L_2^2 - L_3^2, L_1 L_2, L_1 L_3)
    \vspace{2mm} \\
  \vec e_3 = \displaystyle \frac{1}{L}(L_1, L_2, L_3) \mbox{.}
\end{array}
$$

The angular velocity $\omega$ is given by the
equation $\omega^2 = \dot{\vec x}^2 = L^2$, valid to within the
first order of perturbation.
Here $L_1, L_2, L_3$ are coordinates of the normal
to the plane of the great circle determining the solution, i.e.
the angular momentum.

Let us turn to the exact equations for the angular momentum
(\ref{ConcrMom}). With the help of the equations given above and
neglecting terms of the second, and higher, order in the $
\varepsilon_i $, we can transform equations (\ref{ConcrMom}) in
the form

  \begin{equation} 
    \begin{array}{rcl}
    \dot{L_1} = \displaystyle \frac{2 L^2 \varepsilon_2}{(L_2^2 + L_3^2)^2}
       \left[
          \cos(\omega t + \theta) L_3
          + \sin(\omega t + \theta) \displaystyle \frac{L_1 L_2}{L}
       \right]^3
       \left[
           \cos(\omega t + \theta) (-L_2)
           + \sin(\omega t + \theta) \displaystyle \frac{L_1 L_3}{L}
       \right] \vspace{2mm} \\
           - \displaystyle \frac{2 L^2 \varepsilon_3}{(L_2^2 + L_3^2)^2}
       \left[
        \cos(\omega t + \theta) (-L_2)
        + \sin(\omega t + \theta) \displaystyle \frac{L_1 L_3}{L}
       \right]^3
       \left[
         \cos(\omega t + \theta) L_3
          + \sin(\omega t + \theta) \displaystyle \frac{L_1 L_2}{L}
       \right] \\
    \end{array}
    \label{explicit}
  \end{equation}

 $$ 
    \begin{array}{rcl}
       \dot{L_2} = \displaystyle \frac{2 L^2 \varepsilon_3}{(L_2^2 + L_3^2)^2}
       \left[
          \cos(\omega t + \theta) (-L_2)
          + \sin(\omega t + \theta) \displaystyle \frac{L_1 L_3}{L}
       \right]^3
           \sin(\omega t + \theta) \displaystyle \frac{- L_2^2 - L_3^2}{L}
                          \vspace{2mm} \\
          - \displaystyle \frac{2 L^2 \varepsilon_1}{(L_2^2 + L_3^2)^2}
       \left[
              \sin(\omega t + \theta) \displaystyle \frac{- L_2^2 - L_3^2}{L}
       \right]^3
       \left[
         \cos(\omega t + \theta) (-L_2)
        + \sin(\omega t + \theta) \displaystyle \frac{L_1 L_3}{L}
       \right] \\
    \end{array}
 $$

 $$ 
    \begin{array}{rcl}
      \dot{L_3} = \displaystyle \frac{2 L^2 \varepsilon_1}{(L_2^2 + L_3^2)^2}
      \left[
             \sin(\omega t + \theta) \displaystyle \frac{- L_2^2 - L_3^2}{L}
      \right]^3
      \left[
        \cos(\omega t + \theta) L_3
          + \sin(\omega t + \theta) \displaystyle \frac{L_1 L_2}{L}
      \right] \vspace{2mm} \\
          - \displaystyle \frac{2 L^2 \varepsilon_2}{(L_2^2 + L_3^2)^2}
      \left[
        \cos(\omega t + \theta) L_3
          + \sin(\omega t + \theta) \displaystyle \frac{L_1 L_2}{L}
     \right]^3
                 \sin(\omega t + \theta) \displaystyle \frac{- L_2^2 - L_3^2}{L}
      \\
    \end{array}
  $$

It should be noted that the right-hand sides of equations
(\ref{explicit}) comprise terms oscillating in time and terms that
vary slowly. By using the averaging method, \cite{H}, that is on
neglecting the oscillatory terms we obtain the averaged equations
for the angular momentum

  \begin{equation}
    \begin{array}{rcl}
    \displaystyle
    \dot{L_1}
    &=& \displaystyle \frac34 \, \frac{L_2 L_3}{L^2}
    \left[ (\varepsilon_3 - \varepsilon_2) L_1^2 + \varepsilon_3 L_2^2
    - \varepsilon_2 L_3^2
    \right]    \mbox{,}
    \vspace{2mm}    \vspace{2mm} \\

    \displaystyle
    \dot{L_2}
    &=& \displaystyle \frac34 \, \frac{L_3 L_1}{L^2}
    \left[ - \varepsilon_3 L_1^2 + (\varepsilon_1 - \varepsilon_3) L_2^2
    + \varepsilon_1 L_3^2
    \right]    \mbox{,}
    \vspace{2mm}    \vspace{2mm} \\

    \displaystyle
    \dot{L_3}
    &=& \displaystyle \frac34 \, \frac{L_1 L_2}{L^2}
    \left[\varepsilon_2 L_1^2 - \varepsilon_1 L_2^2 +
    (\varepsilon_2 - \varepsilon_1) L_3^2
    \right]    \mbox{.}
    \vspace{2mm}            \\
    \end{array}
    \label{avmom}
  \end{equation}

It is worth noting that equations (\ref{avmom}) have the
Hamiltonian form determined by the usual Poisson brackets for the
angular momentum, \cite{R},
$$
    \{L_i, L_j\} = \sum_k \varepsilon_{ijk} L_k  \mbox{,}
$$
and the Hamiltonian

  \begin{equation}
    H = \frac{3}{16} L^2 \sum_i \varepsilon_i
             \left[ \left( \frac{L_i}{L} \right)^2 - 1
             \right]^2  \mbox{.}
    \label{hamiltonian}
  \end{equation}

\noindent This circumstance is particularly interesting because,
usually, the averaging procedure is not compatible with
Hamiltonian structure. The system we have obtained is the
integrable Hamiltonian one, but its exact solution is cumbersome.
Therefore, we shall find a qualitative description of the system's
motion and extensively use numerical simulation.

The important point is considering the stationary solutions to
equations (\ref{avmom}) for which the right-hand sides turn out to
be zero. They correspond to trajectories that are closed to within
oscillating terms neglected during the averaging.  Their equations
split into three parts {\sf S1, S2} and {\sf S3}, determined by
conditions on $\varepsilon_i$, as follows.

\begin{itemize}
  \item[\sf S1] \label{S1}
    No algebraic constraints imposed on $\varepsilon_i$:
    \begin{itemize}
    \item[\sf a.] \label{T1a}
    $ L_{10}  =  0, \quad L_{20}  =  0, \quad L_{30} \ne 0; $
    \item[\sf b.] \label{T1b}
        $ L_{10}  =  0, \quad L_{20} \ne 0, \quad L_{30}  =  0; $
        \item[\sf c.] \label{T1c}
     $ L_{10} \ne 0, \quad L_{20}  =  0, \quad L_{30}  =  0; $
    \end{itemize}
  \item[\sf S2] \label{S2}
    The constraints on $\vec L$ relaxed and linear constraints
    imposed on $\varepsilon_i$:
    \begin{itemize}
    \item[\sf a.] \label{T2a}
    $ L_{10} = 0, \quad L_{20} \ne 0, \quad L_{30} \ne 0, \quad
      \varepsilon_3 L_{20}^2 - \varepsilon_2 L_{30}^2 = 0; $
    \item[\sf b.] \label{T2b}
      $ L_{20} = 0, \quad L_{30} \ne 0, \quad L_{10} \ne 0, \quad
       \varepsilon_1 L_{30}^2 - \varepsilon_3 L_{10}^2 = 0; $
    \item[\sf c.] \label{T2c}
     $L_{30} = 0, \quad L_{10} \ne 0, \quad L_{20} \ne 0, \quad
     \varepsilon_2 L_{10}^2 - \varepsilon_1 L_{20}^2 = 0; $
    \end{itemize}
  \item[\sf S3] \label{S3}
    Vector $\vec L$ subject to  $L_{10} \ne 0, L_{20} \ne 0,  L_{30} \ne 0$
    and the quadratic constraints imposed on
    $\varepsilon_i$:
    $$
        \frac{L_{10}^2}
        {\varepsilon_1 \,\varepsilon_2 - \varepsilon_2 \,
        \varepsilon_3 + \varepsilon_3
        \,\varepsilon_1} =
        \frac{L_{20}^2}
        {\varepsilon_1 \,\varepsilon_2 + \varepsilon_2
        \,\varepsilon_3 - \varepsilon_3
        \,\varepsilon_1} =
        \frac{L_{30}^2}
        { - \varepsilon_1 \,\varepsilon_2 + \varepsilon_2
        \,\varepsilon_3 + \varepsilon_3
        \,\varepsilon_1}
    $$
 \end{itemize}

It is worth noting that equations  {\sf S2} involve the fulfilment
of the inequalities $\varepsilon_2 \varepsilon_3 > 0$,
$\varepsilon_3 \varepsilon_1 > 0$, and $\varepsilon_1
\varepsilon_2 > 0$ for cases {S2.a, S2.b, S2.c}, respectively,
whereas equations {\sf S3} involve
        \begin{eqnarray}
            \varepsilon_1 \,\varepsilon_2 - \varepsilon_2 \,\varepsilon_3 +
            \varepsilon_3
            \,\varepsilon_1  \, > 0  \nonumber  \\
            \varepsilon_1 \,\varepsilon_2 + \varepsilon_2 \,\varepsilon_3 -
            \varepsilon_3
            \,\varepsilon_1  \, > 0  \nonumber  \\
            \ - \varepsilon_1 \,\varepsilon_2 + \varepsilon_2 \,\varepsilon_3
            + \varepsilon_3
            \,\varepsilon_1  \, > 0  \nonumber
        \end{eqnarray}

Linearizing equations (\ref{avmom}) at the stationary solutions and,
considering small fluctuations of $\vec L$ round them, we may
study their stability, which turns out to be determined by the
requirements
\begin{itemize}
    \item[\sf S1]
    \begin{itemize}
        \item[\sf a.] $\varepsilon_1 \varepsilon_2 > 0$;
        \item[\sf b.] $\varepsilon_2 \varepsilon_3 > 0$;
        \item[\sf c.] $\varepsilon_3 \varepsilon_1 > 0$.
    \end{itemize}
    \item[\sf S2]
    \begin{itemize}
        \item[\sf a.] $ \varepsilon_1 \varepsilon_2
            -\varepsilon_2 \varepsilon_3
            +\varepsilon_3 \varepsilon_1 < 0 $;
        \item[\sf b.] $ \varepsilon_1 \varepsilon_2
            +\varepsilon_2  \varepsilon_3
            -\varepsilon_3 \varepsilon_1 < 0 $;
        \item[\sf c.] $-\varepsilon_1 \varepsilon_2
            +\varepsilon_2 \varepsilon_3
            +\varepsilon_3 \varepsilon_1 < 0 $.
    \end{itemize}
    \item[\sf S3] \quad any $ \varepsilon_i $.
\end{itemize}

We may  put these equations in a more graphic form by using the
integral $ L^2 = const $, and consider the motion of $\vec L$ on a
sphere of  fixed radius, the integral of energy $H$ taking
appropriate values. Then the stable solutions are fixed points as
regards equations (\ref{avmom}),  the stable and the unstable
points are foci and saddle points, respectively, the separaterixes
being lines joining the fixed points. Together, they generate a
graph on the sphere, having the fixed points as vertices and the
separaterixes as edges.  It is important that the separaterixes,
i.e. the edges of the graph, are oriented according to  time, $t$,
so that the graph is the oriented one, and invariant with respect
to the symmetry $ \vec R \rightarrow - \vec R $ and time inversion
$ t \rightarrow - t $.

\begin{figure}
  \begin{center}
    \includegraphics[width = 200bp]{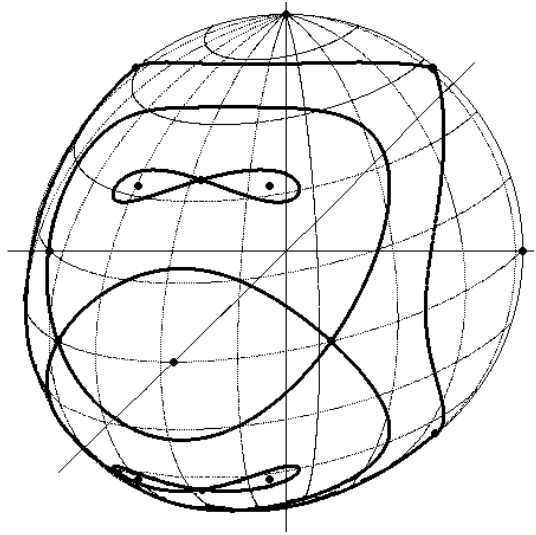}
    \caption{Separatrix net corresponding to the graph of
     {\sf Type I}  on the sphere.
    }
    \label{fig2}
  \end{center}
\end{figure}

\begin{figure}
  \begin{center}
    \includegraphics[width = 200bp]{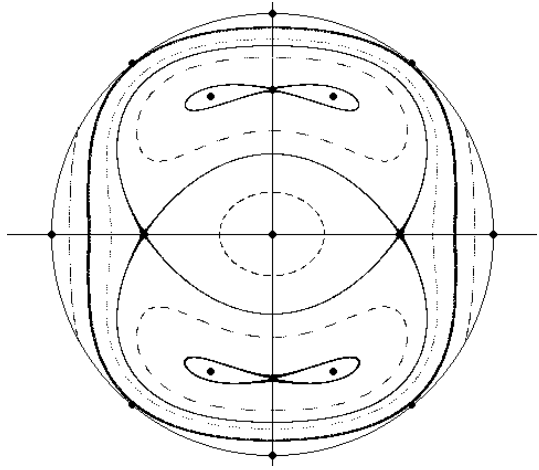}
    \caption{ {\sf Type I} trajectories of the auxiliary system on the
      semi-sphere; dashed lines are typical trajectories,
      the solid ones separatrixes.
      Parameters of deformation:
      $ \varepsilon_1 = 0.02, \
    \varepsilon_2 = 0.03, \
    \varepsilon_3 = 0.04. $
    }
    \label{fig3}
  \end{center}
\end{figure}

\begin{figure}
  \begin{center}
    \includegraphics[width = 200bp]{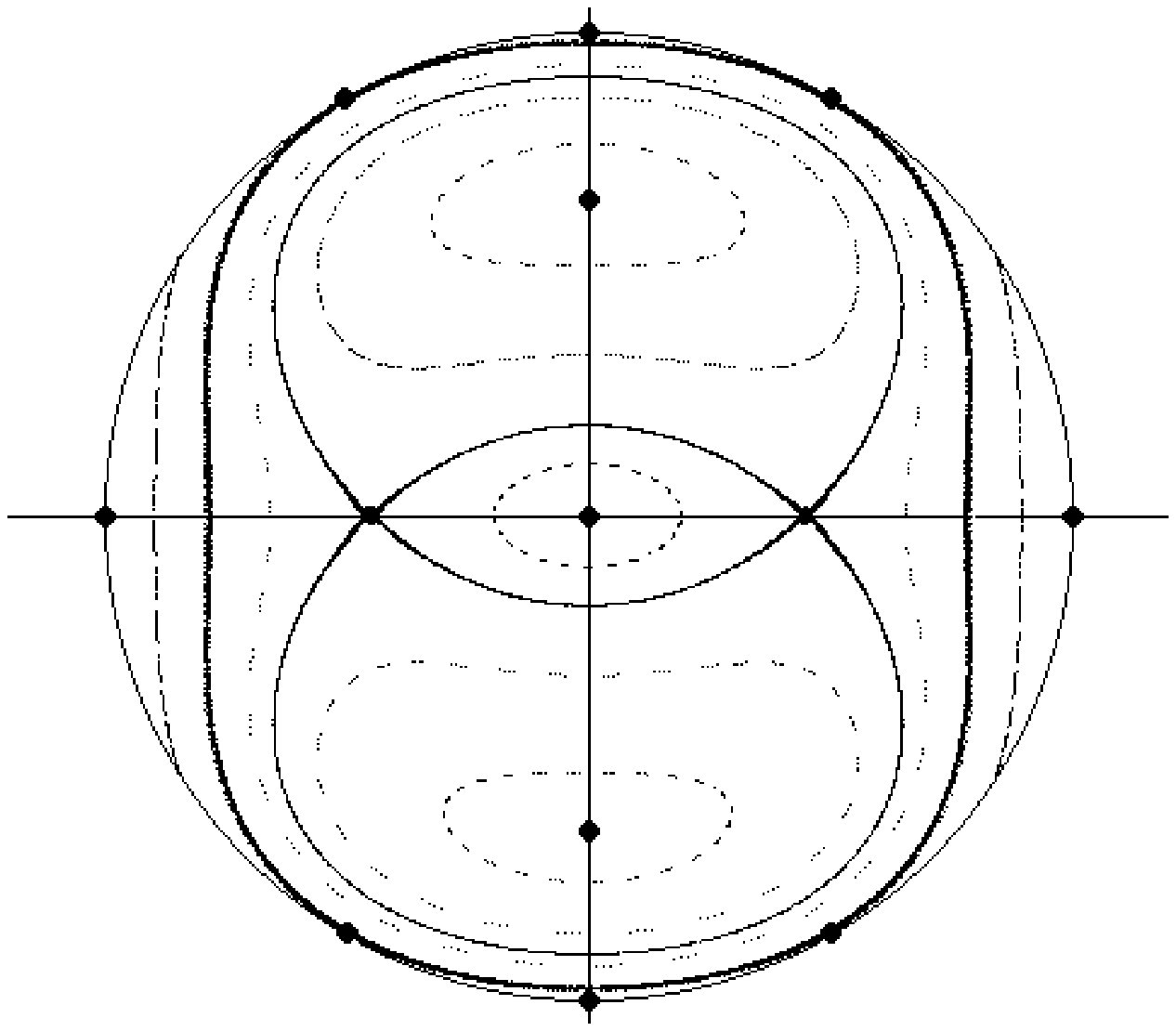}
    \caption{{\sf Type II} trajectories of the auxiliary system on the
      semi-sphere; dashed lines are typical trajectories,
      the solid ones separatrixes.
      Parameters of deformation:
      $ \varepsilon_1 = 0.01, \
    \varepsilon_2 = 0.03, \
    \varepsilon_3 = 0.04. $
    }
    \label{fig4}
  \end{center}
\end{figure}

\begin{figure}
  \begin{center}
    \includegraphics[width = 200bp]{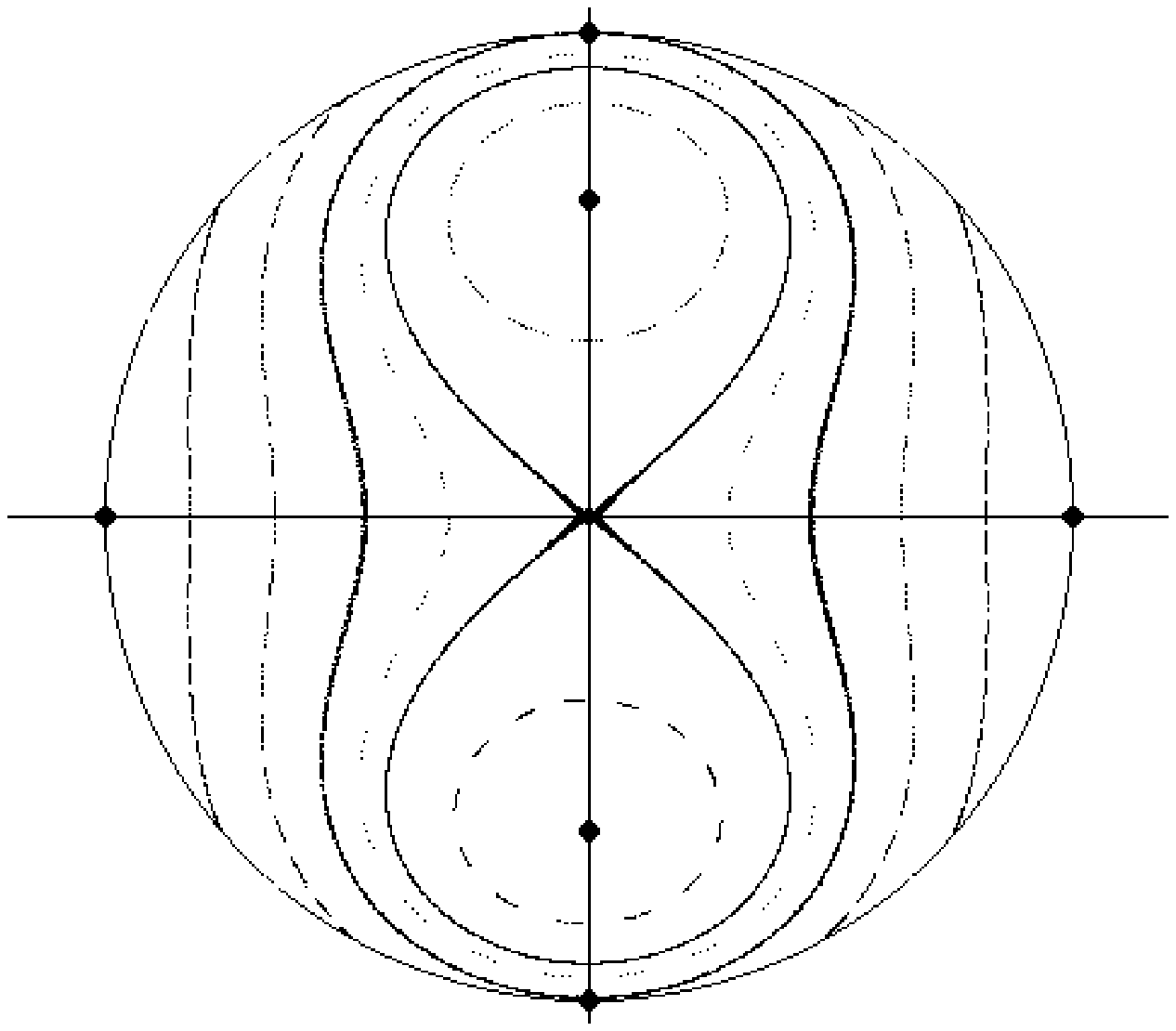}
    \caption{{\sf Type III} trajectories of the auxiliary system on the
      semi-sphere; dashed lines are typical trajectories,
      the solid ones separatrixes.
      Parameters of deformation:
      $ \varepsilon_1 = -0.02, \
    \varepsilon_2 = 0.03, \
    \varepsilon_3 = 0.04. $
    }    \label{fig5}
  \end{center}
\end{figure}

\begin{figure}
  \begin{center}
    \includegraphics[width = 200bp]{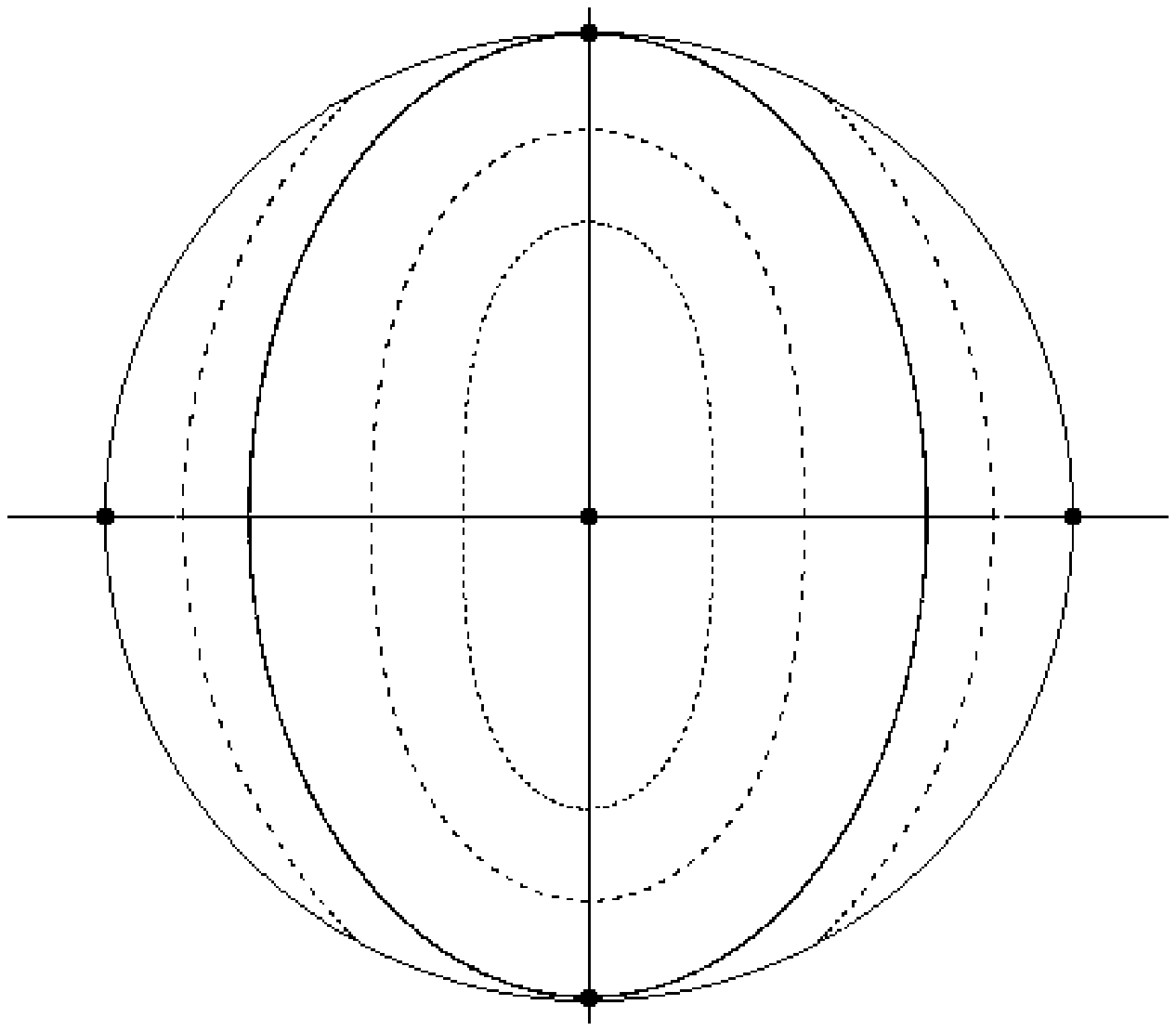}
    \caption{{\sf Type IV} trajectories of the auxiliary system on the
      semi-sphere; dashed lines are typical trajectories,
      the solid ones separatrixes.
      Parameters of deformation:
      $ \varepsilon_1 = -0.01, \
    \varepsilon_2 = 0.00, \
    \varepsilon_3 = 0.01. $
     }
     \label{fig6}
 \end{center}
\end{figure}

Now we are in a position to determine the graphs by employing the
computer simulation of the equations (\ref{avmom}), and using the
types of the fixed points found above. It should be noted that we
must check as to whether the solutions provided by equations
(\ref{avmom}) agree with those given by original equations
(\ref{2Newton}), see FIG.\ref{fig8}. The phase picture can be
obtained by constructing a mesh generated by solutions to
equations (\ref{avmom}), taking into account the types of fixed
points. The results are illustrated in
FIGs.\ref{fig3}--\ref{fig6}.

We obtain the following  types of the graphs:

\begin{itemize}
  \item[\sf Type I] \label{T1}
        \quad FIG.\ref{fig3},
        \quad  7 foci and 6 saddles;
        $ \varepsilon_i $ being subject to the constraints:
      \begin{equation}
      \begin{array}{rcl}
     \varepsilon_1 \varepsilon_2
    -\varepsilon_2 \varepsilon_3
    +\varepsilon_3 \varepsilon_1 > 0 \\

     \varepsilon_1 \varepsilon_2
    +\varepsilon_2 \varepsilon_3
    -\varepsilon_3 \varepsilon_1 > 0 \\

    -\varepsilon_1 \varepsilon_2
    +\varepsilon_2 \varepsilon_3
    +\varepsilon_3 \varepsilon_1 > 0
      \end{array}
      \mbox{.}
     \label{t1constraints}
      \end{equation}

 \item[\sf Type II] \label{T2}
       \quad FIG.\ref{fig4},
       \quad 5 foci and 4 saddle points;
       $ \varepsilon_i $ are not equal to zero,
       have the same sign,
       and at least one of
       equations (\ref{t1constraints}) is not true.

 \item[\sf Type III] \label{T3}
       \quad FIG.\ref{fig5},
       \quad 3 foci and 2 saddle points;
       $ \varepsilon_i $ being subject to
       one of the following constraints:
       $ \varepsilon_2 \varepsilon_3 > 0 $ and
       $ \varepsilon_1 \varepsilon_2 \leq 0 $; \quad
       $ \varepsilon_3 \varepsilon_1 > 0 $ and
       $ \varepsilon_2 \varepsilon_3 \leq 0 $; \quad
       $ \varepsilon_1 \varepsilon_2 > 0  $ and
       $ \varepsilon_3 \varepsilon_1 \leq 0 $.

 \item[\sf Type IV] \label{T4}
      \quad FIG.\ref{fig6},
      \quad 2 foci and 1 saddle point;
      $ \varepsilon_i $ being subject to
      one of the following constraints:
       $ \varepsilon_1 = 0 $ and
       $ \varepsilon_2 \varepsilon_3 \leq 0 $; \quad
       $ \varepsilon_2 = 0 $ and
       $ \varepsilon_3 \varepsilon_1 \leq 0 $; \quad
       $ \varepsilon_3 = 0 $ and
       $ \varepsilon_1 \varepsilon_2 \leq 0 $.

  \end{itemize}

\begin{figure}
  \begin{center}
    \includegraphics[width = 400bp]{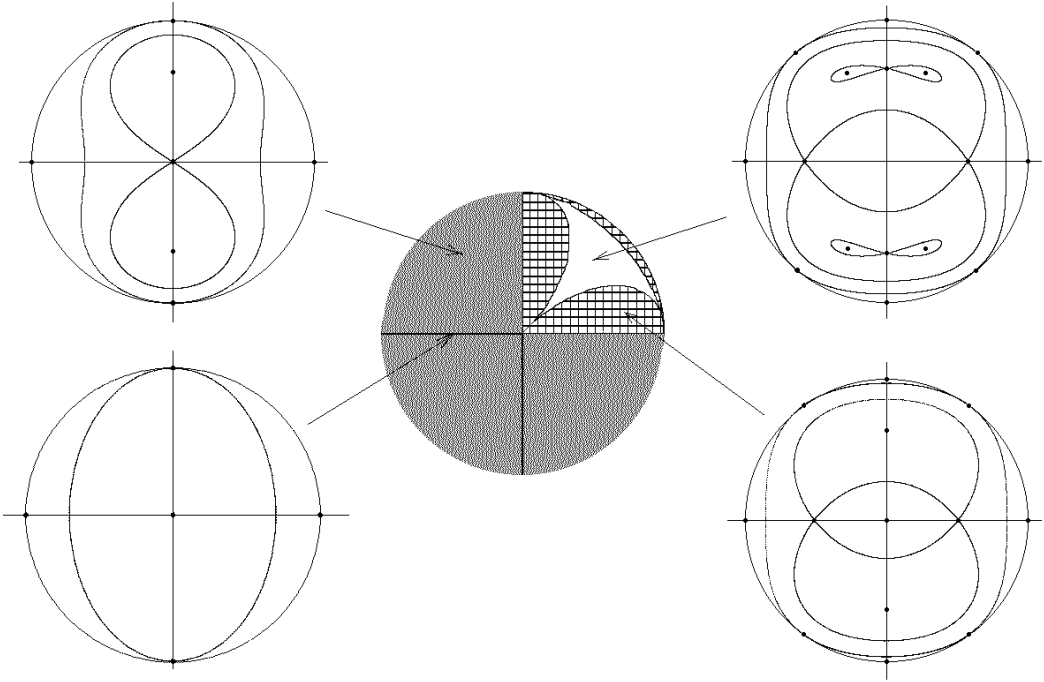}
    \caption{
     Regions of $\varepsilon_i$ corresponding to Types I - IV of
     the solutions to the auxiliary system. The white area
     indicates {\sf Type I} solutions, the filled and the
     barred ones {\sf Type II} and {\sf Type III}.
     The lines dividing the {\sf Type I} and {\sf Type II} regions
     are subject to equations (\ref{boundary})
    }
    \label{fig7}
  \end{center}
\end{figure}

\begin{figure}
  \begin{center}
    \includegraphics[width = 400bp]{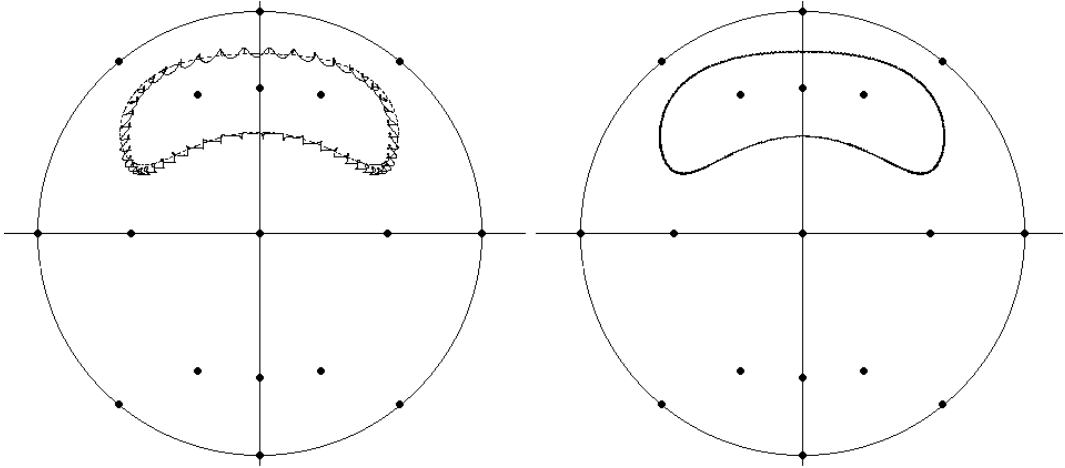}
    \caption{Comparison of  solution to: A. the initial equations
     for geodesics; B. the averaged equation given by the
     auxiliary system.
    }
    \label{fig8}
  \end{center}
\end{figure}

The separatrix nets depend on values of the coefficients of the
deformation $\varepsilon_i$, and generate regions I, II, III, IV
in the $\varepsilon_i$ space.

It is important that the lines dividing the domains corresponding
to types I and II, FIG.\ref{fig7}, are given by the homogeneous equations

\begin{eqnarray}
   &1.&\quad \varepsilon_1 \varepsilon_2
  -\varepsilon_2 \varepsilon_3
  +\varepsilon_3 \varepsilon_1 = 0 \label{boundary} \\
   &2.&\quad \varepsilon_1 \varepsilon_2
  +\varepsilon_2 \varepsilon_3
  -\varepsilon_3 \varepsilon_1 = 0 \nonumber \\
   &3.&\quad -\varepsilon_1 \varepsilon_2
  +\varepsilon_2 \varepsilon_3
  +\varepsilon_3 \varepsilon_1 = 0 \nonumber
\end{eqnarray}

The type of a graph corresponding to the solutions  is completely
determined by the numbers of foci and saddle points. The
dependence of the conformations of the foci and the saddles on
values of $\varepsilon_i$ is illustrated in FIG.\ref{fig7}.

   \section{Conclusion}
    \label{conclusion}

The main instrument  of the present investigation is the auxiliary
Hamiltonian system, which can be considered as a reduction of the
initial problem to a dynamical problem on specific configuration
and phase spaces. Points of the new configuration space are
geometrical objects, i.e great circles, of the configuration
space, i.e. the standard sphere. Thus, we obtain a Hamiltonian
system that describes the transformation of these objects. In a
sense, the approach  follows the classical method of Klein and
Lie, \cite{Klein}, of constructing a new space with objects of the
given one. For the specific case of an ellipsoid close to the
standard sphere our classification of orbits are in agreement with
the classical results by Jacobi, \cite{J} for geodesics on
ellipsoid. We feel that our asymptotic approach is useful for the
treatment of more general problems, for example the motion of
rigid bodies, and intend to consider the problem in the subsequent
paper. In analytical terms we study an asymptotic reduction of the
system of equations for orbits on a deformed sphere to that of the
top, but with the Hamiltonian of the fourth order. The
simplification we get in this way, is substantial. Indeed, the
initial Hamiltonian system  could be non-integrable,whereas the
auxiliary system is totally integrable and described by a graph
that comprises vertices, which correspond to stationary solutions,
or almost closed orbits, and edges, which can be visualized as
orbits joining them, that is orbits which continuously approach
more and more closely to coincidence with the closed ones.

\end{document}